\documentclass[graybox]{svmult}
\usepackage{mathptmx}       
\usepackage{helvet}         
\usepackage{courier}        
\usepackage{makeidx}         
\usepackage{graphicx}        
\usepackage{multicol}        
\usepackage[bottom]{footmisc}
\usepackage{amsmath,amssymb,amsfonts}        
\usepackage{tikz}
\usepackage{doi}
\makeindex             

\begin{document}
\title*{Superintegrable bosonic star networks}
\author{Angela Foerster and Jon Links}
\institute{Angela Foerster \at Instituto de F\'{i}sica da UFRGS, Av. Bento Gon\c{c}alves 9500, Agronomia, Porto Alegre, RS 91501-970, Brazil, \email{angela@if.ufrgs.br}
\and Jon Links \at School of Mathematics and Physics, The University of Queensland, 4072, Australia, \\ \email{jrl@maths.uq.edu.au}}
%
%
\maketitle

\abstract{We introduce a class of bosonic star networks involving a central site interacting with the surrounding environment sites. These networks are shown to be {\it superintegrable}. We present two forms of Bethe Ansatz solution providing expressions for the energy eigenvalues. A brief discussion is included on the potential applications.  
}

\section{Introduction}
\label{sec:1}
There has recently been significant interest in quantum models describing a central degree of freedom interacting with an environment modeled as an ensemble of surrounding degrees of freedom. This has been especially the case for spin degrees of freedom, giving rise to the class of {\it central spin models}, also known as {\it spin stars}. Many of these examples are integrable systems admitting Bethe Ansatz solutions, allowing for detailed analyses of their properties to be undertaken e.g. \cite{df22,ng18,vl23,vcc20}.   

In this note we provide an analogous class of models for bosonic systems, consisting of a single bosonic site coupled to environment sites through tunneling terms, as well as a global collective interaction. These models are defined in Sect. \ref{sec:2}. Our interest in these is that they are superintegrable in the sense that there exists a larger number of conserved operators than degrees of freedom. We will explicitly construct the conserved operators in Sect. \ref{sec:3}. Identifying these facilitates the calculation of a Bethe Ansatz solution for determining the energy spectrum. Results on this aspect are collated in Sect \ref{sec:4}. Concluding remarks are provided in Sect. \ref{sec:5}.  

\section{The models}
\label{sec:2}

We introduce the star model Hamiltonian $H_{(p,1)}$ for $p+1$ sites in terms of a set of canonical boson operators $a_i,\,a_i^\dagger, N_{i}=a_i^\dagger a_i$, $i=1,...,p$, and another set 
$b,\,b^\dagger, N_b=b^\dagger b$. The Hamiltonian reads
\begin{align}
\label{ham}
H_{(n,1)} 
&=U(N_{a}-N_{b})^2-J\sum_{j=1}^p (a_{j}b^\dagger+a_{j}^\dagger b),
\end{align}
where we have defined 
\begin{align*}
N_{a}= 
\sum_{j=1}^p a^\dagger_{j}a_j.    
\end{align*}
The coupling parameter $U$ provides the interaction strength between the central site and the surrounding environment sites, while the parameter $J$ is the coupling strength for the hopping between sites. These star networks belong to a larger class of networks that are associated with complete bipartite graphs \cite{bfil24}. 
A schematic representation of these systems is provided below in Fig. 1.

\begin{figure}[h] \label{fig:1}
\centering

\begin{tikzpicture}
\node[circle, fill=blue, scale=0.7] (n0) at (1, 0) {};
\node[circle, fill=teal, scale=0.7] (n1) at (0, 0) {};
\node[circle, fill=teal, scale=0.7] (n2) at (2, 0) {} ;
\node[circle, fill=teal, scale=0.7] (n3) at (1, 1) {};
\node[circle, fill=teal, scale=0.7] (n4) at (1, -1) {};
\filldraw[ draw=black, line width=1.5pt]  (n0)--(n1);
\filldraw[ draw=black, line width=1.5pt]  (n0)--(n2);
\filldraw[ draw=black, line width=1.5pt]  (n0)--(n3);
\filldraw[ draw=black, line width=1.5pt]  (n0)--(n4);
            
\node[circle, fill=blue, scale=0.7] (m0) at (4, 0) {};
\node[circle, fill=teal, scale=0.7] (m1) at (4, -1) {};
\node[circle, fill=teal, scale=0.7] (m2) at (4+0.95106, -0.30902) {};
\node[circle, fill=teal, scale=0.7] (m3) at (4-0.95106, -0.30902) {};
\node[circle, fill=teal, scale=0.7] (m4) at (4+0.58779, 0.80897) {};
\node[circle, fill=teal, scale=0.7] (m5) at (4-0.58779, 0.80907) {};
\filldraw[ draw=black, line width=1.5pt]  (m0)--(m1);
\filldraw[ draw=black, line width=1.5pt]  (m0)--(m2);
\filldraw[ draw=black, line width=1.5pt]  (m0)--(m3);
\filldraw[ draw=black, line width=1.5pt]  (m0)--(m4);  
\filldraw[ draw=black, line width=1.5pt]  (m0)--(m5);  

\node[circle, fill=blue, scale=0.7] (l0) at (7, 0) {};
\node[circle, fill=teal, scale=0.7] (l1) at (7, 1) {};
\node[circle, fill=teal, scale=0.7] (l2) at (7, -1) {};
\node[circle, fill=teal, scale=0.7] (l3) at (7+0.86603, 0.5) {};
\node[circle, fill=teal, scale=0.7] (l4) at (7-0.86603, 0.5) {};
\node[circle, fill=teal, scale=0.7] (l5) at (7+0.86603, -0.5) {};
\node[circle, fill=teal, scale=0.7] (l6) at (7-0.86603, -0.5) {};
\filldraw[ draw=black, line width=1.5pt]  (l0)--(l1);
\filldraw[ draw=black, line width=1.5pt]  (l0)--(l2);
\filldraw[ draw=black, line width=1.5pt]  (l0)--(l3);
\filldraw[ draw=black, line width=1.5pt]  (l0)--(l4);  
\filldraw[ draw=black, line width=1.5pt]  (l0)--(l5);
\filldraw[ draw=black, line width=1.5pt]  (l0)--(l6);

\node[circle, fill=blue, scale=0.7] (p0) at (10, 0) {};
\node[circle, fill=teal, scale=0.7] (p1) at (10, -1) {};
\node[circle, fill=teal, scale=0.7] (p2) at (10+0.97493, 0.22252) {};
\node[circle, fill=teal, scale=0.7] (p3) at (10-0.97493, 0.22252) {};
\node[circle, fill=teal, scale=0.7] (p4) at (10-0.43388, 0.90097) {};
\node[circle, fill=teal, scale=0.7] (p5) at (10+0.43388, 0.90097) {};
\node[circle, fill=teal, scale=0.7] (p6) at (10+0.78183, -0.62349) {};
\node[circle, fill=teal, scale=0.7] (p7) at (10-0.78183, -0.62349) {};
\filldraw[ draw=black, line width=1.5pt]  (p0)--(p1);
\filldraw[ draw=black, line width=1.5pt]  (p0)--(p2);
\filldraw[ draw=black, line width=1.5pt]  (p0)--(p3);
\filldraw[ draw=black, line width=1.5pt]  (p0)--(p4);  
\filldraw[ draw=black, line width=1.5pt]  (p0)--(p5);  
\filldraw[ draw=black, line width=1.5pt]  (p0)--(p6);  
\filldraw[ draw=black, line width=1.5pt]  (p0)--(p7);

        \end{tikzpicture}
\caption{Pictorial representation of the star network models $H_{(4,1)}$ on five sites (leftmost) through to $H_{(7,1)}$ on eight sites (rightmost). }
\end{figure}
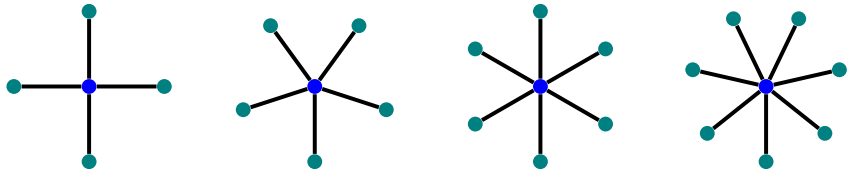
The fundamental case of the $H_{(2,1)}$ Hamiltonian was first introduced in \cite{wytlf18} as a model of cold dipolar bosonic atoms confined in a triple well potential. In that work,  an analysis was undertaken to gain insights into the functionality of the system as an atomtronic switch. This same model has been subsequently studied in relation to generating and controlling entanglement \cite{tywfl20,wyblf23}, as well as the transition to quantum chaos under the breaking of integrability through the inclusion of additional interactions terms \cite{ccrsh21,cwcrfh24,wcfs22}. 

\section{Superintegrability}
\label{sec:3}  

It is easily verified that the Hamiltonian (\ref{ham}) commutes with the total particle number operator given by 
$N=N_{a}+N_{b}$. In addition, the Hamiltonian commutes with the generators for a realisation of $gl(p-1)$. This result shows that for $p\geq 3$ the Hamiltonian is superintegrable - the number of degrees of freedom in the system, which is $p+1$, is less than the number of algebraically independent conserved operators, which is $2p-1$.

To identify superintegrability, a discrete Fourier transform is implemented via
\begin{align*}
c_k=\frac{1}{\sqrt{p}}\sum_{j=1}^p \exp\left(\frac{2\pi ij(k-p)}{p}  \right) a_j   , \qquad  c^\dagger_k=\frac{1}{\sqrt{p}}\sum_{j=1}^p \exp\left(\frac{-2\pi ij(k-p)}{p}  \right) a^\dagger_j,
\end{align*}
such that the canonical bosonic commutation relations are preserved and 
\begin{align*}
N_c  = \sum_{j=1}^p c^\dagger_{j}c_j = N_a 
\end{align*} 
The Hamiltonian (\ref{ham}) then assumes the form
\begin{align}
\label{tham}
H_{(n,1)} 
&=U(N_{b}-N_{c})^2-J\sqrt{p}(b^\dagger c_p+c_p^\dagger b).
\end{align}
In this form it becomes apparent that the operators $\{E^j_k=c_j^\dagger c_k: k=1,\dots,p-1\}$ satisfy the $gl(p-1)$ commutation relations 
\begin{align*}
[E^j_k,\,E^m_n]=\delta^m_k E^j_n-\delta^j_n E^m_k    
\end{align*}
and commute with the Hamiltonian (\ref{tham}), demonstrating superintegrability. 

Moreover, the states 
\begin{align}
|r_1,\dots,r_{p-1}\rangle=  \prod_{k=1}^{p-1} (c_k^\dagger)^{r_k} |0\rangle ,     
\label{pvac} 
\end{align}
where $|0\rangle$ is the Fock vacuum, are eigenstates of (\ref{tham}). The corresponding eigenvalues are given by $Ur^2$, where $\displaystyle r=\sum_{k=1}^{p-1} r_k$.
The states (\ref{pvac}) serve as pseudovacuum states in the derivation of the Bethe Ansatz results stated below.

\section{Bethe Ansatz solutions}
\label{sec:4}
There are two known forms of Bethe Ansatz solution for the above star Hamiltonians. One is a specialisation of results derived in 2017 \cite{ytfl17} through the standard formulation of the Algebraic Bethe Ansatz (ABA). Later, in 2024, an alternative form was obtained \cite{bfil24} via a different means. In the first instance the Bethe Ansatz Equations (BAE) have a multiplicative form, and are amenable for studies in the weak interaction regime following the approach of \cite{zlmg03}. In the second case the BAE are additive, and are amenable for studies in the strong interaction regime as described in \cite{bfil24}. 

Set $\displaystyle \eta^2=\frac{4U}{J\sqrt{p}}$. Using the ABA,
the energies of the Hamiltonian (\ref{ham}) are provided via solutions of the Bethe Ansatz equations
\begin{align}
\label{bae}
\eta^2 v_i\left(v_i+\eta r\right) &= \prod_{j\neq i}^{N-r}\frac{v_i-v_j-\eta}{v_i-v_j+\eta}, \quad r<N.
\end{align} 
Given a solution to (\ref{bae}), the associated energy reads
\begin{align}
\label{h2ne}
E=-J\sqrt{p} \left(\lambda_{r}(u)-u^2-\eta^{-2}-u \eta N-\frac{\eta^2N^2}{4}\right),
\end{align}
where 
\begin{align*}
\lambda_{r}(u) &= u\left(u+\eta r \right) \prod_{j=1}^{N-r}\frac{u-v_j+\eta}{u-v_j}+\eta^{-2}\prod_{j=1}^{N-r}\frac{u-v_j-\eta}{u-v_j}. 
\end{align*} 
Note that for $r=N$ the above is to be interpreted as  $ \lambda_N (u) =u\left(u+\eta r\right) +\eta^{-2}$ such that $E=Ur^2$.

The alternative exact-solution formulation developed in \cite{bfil24} yields the energy expression 

\begin{align}
E = U(N-2r)^{2} - J\sqrt{p}\sum^{N-r}_{j=1}u_{j},
\label{nrg2}
\end{align}
where $\{ u_{n} : n=1,\ldots, N-r\}$ is the set of solutions to the BAE
\begin{align}
\label{bae2}
    \dfrac{J\sqrt{p}}{4U}u_{k}^{-2} + (1 - N)u_{k}^{-1} - \dfrac{J\sqrt{p}}{4U} = \sum^{N-r}_{j\neq k}\dfrac{2}{u_{j} - u_{k}}.
\end{align}
We remark that, using (\ref{bae2}), the formula (\ref{nrg2}) can also be expressed as 
\begin{align}
E= UN^{2} - J\sqrt{p}\sum^{N-r}_{j=1}u_{j}^{-1}.
\label{nrg3}
\end{align}

\section{Discussion}
\label{sec:5}
We have introduced a class of bosonic superintegrable star networks
and presented two distinct Bethe Ansatz solutions.
In future work we will explore
potential applications of these networks.
Starting with the 3+1 model, we have found that this provides a framework for
the design of a directional quantum switch
and a diplexer-like atomic system.

Due to the dihedral symmetry of the star networks, 
a relevant question is:
Can we enhance the design of quantum devices by rotating these systems?
We have found that the superintegrability property
plays a crucial role in this regard, which will be reported in upcoming work.

\begin{acknowledgement}
This research was supported by the Australian Research Council through Discovery Project DP200101339. AF also acknowledges support from  
CNPq (Conselho Nacional de Desenvolvimento Científico e Tecnológico) - Edital Universal 406563/2021-7. 
We thank the mathematical research institute MATRIX in Australia where part of this research was performed.
JL acknowledges the traditional owners of the land on which The University of Queensland (St. Lucia campus) operates, the Turrbal and Jagera people
\end{acknowledgement}

\end{document}